\documentstyle[pra,aps,preprint]{revtex}
\include{epsf}
\begin{document}
\tightenlines
\title{Theory of Four Wave Mixing of Matter Waves from a Bose-Einstein
Condensate}

\author{Marek Trippenbach$^{\,1}$, Y.\ B.\ Band$^{\,1}$ and P.\ S.\
Julienne$^{\,2}$}

\address{${}^{1}$ Departments of Chemistry and Physics, \\ Ben-Gurion
University of the Negev, Beer-Sheva, Israel 84105 \\ ${}^{2}$ Atomic
Physics Division, A267 Physics\\ National Institute of Standards and
Technology, Gaithersburg, MD  20899}

\maketitle

\begin{abstract} A recent experiment [Deng et al., Nature 398, 218(1999)]
demonstrated four-wave mixing of matter wavepackets created from
a Bose-Einstein condensate.  The experiment utilized light pulses to
create two high-momentum wavepackets via Bragg diffraction from a
stationary Bose-Einstein condensate.  The high-momentum components
and the initial low momentum condensate interact to form a new
momentum component due to the nonlinear self-interaction of the
bosonic atoms.  We develop a three-dimensional quantum mechanical
description, based on the slowly-varying-envelope approximation, for
four-wave mixing in Bose-Einstein condensates using the time-dependent
Gross-Pitaevskii equation.  We apply this description to describe the
experimental observations and to make predictions.  We examine the role of
phase-modulation, momentum and energy conservation (i.e., phase-matching),
and particle number conservation in four-wave mixing of matter waves,
and develop simple models for understanding our numerical results.
\end{abstract}


\section{Introduction}

Nonlinear optics has been made possible by the nonlinear nature of the
interaction between light and matter and by the development of intense
light sources that can probe the nonlinear regime of this interaction. 
Nonlinear optical processes include three- and four-wave mixing (4WM)
processes (e.g., second harmonic generation and third harmonic
generation).  In 4WM three waves (or light pulses) mix to produce a
fourth.  In this paper we detail our studies of 4WM of coherent matter
waves.  Trippenbach {\it et al.}~\cite{Tripp98} proposed a 4WM
experiment using three colliding Bose-Einstein condensate (BEC)
wavepackets with different momenta.  Deng {\it et al.}~\cite{Deng}
successfully demonstrated 4WM in an experiment with three BEC
wavepackets, which interact in a nonlinear manner to make a fourth BEC
wavepacket.  Here we greatly elaborate on and further develop the
theory and describe numerical simulations of the 4WM output that agree
well with the experimental measurements of ~\cite{Deng}.

The experimental study of nonlinear atom optics is made possible by
the advent of Bose-Einstein condensation of dilute atomic
gases~\cite{atom_BECs,reviews} and the atom ``laser''~\cite{Mewesoc},
a source of coherent matter-waves analogous to the output of optical
lasers.  A set of optical light pulses incident on a parent condensate
with momentum ${\bf P}_1 = {\bf 0}$ can, by Bragg
scattering~\cite{Kozuma99}, create two new daughter BEC wavepackets
with momenta ${\bf P}_{2}$ and ${\bf P}_{3}$.  Four-wave mixing in a
single spin-component condensate occurs as a result of the nonlinear
self-interaction term in the Hamiltonian for a BEC when three such BEC
wavepackets with momenta ${\bf P}_1$, ${\bf P}_2$, and ${\bf P}_3$
collide and interact.  The nonlinear self-interaction can generate a
new BEC wavepacket with a new momentum ${\bf P}_4 = {\bf P}_1 - {\bf
P}_2 + {\bf P}_3$.

The possibility of nonlinear effects in atom optics has been long
recognized~\cite{Lenz}.  Goldstein {\it et al.}~\cite{Goldstein95}
proposed that phase conjugation of matter waves should be possible in
analogy to this phenomenon in nonlinear optics, including the case of
multiple spin-component condensates~\cite{Goldstein99}.  They
considered the case where a ``probe'' BEC wavepacket interacts with
two counter-propagating ``pump'' wavepackets to generate a fourth that
is phase conjugate to the probe, where the probe is weak and causes
negligible depletion of the pump.  Law {\it et al.}~\cite{Law98} also
suggested analogies between interactions in multiple spin-component
condensates and four-wave mixing.  Goldstein and
Meystre~\cite{Goldstein99b} develop a theory of 4WM in multicomponent
BECs based on an algebraic angular momentum approach to obtain the
modes of the coupled operator equations.  Our treatment for a single
spin-component condensate is based on the time-dependent
Gross-Pitaevskii equation (GPE), which has proved to be highly
successful in describing the properties of a variety of actual BEC
experiments~\cite{reviews}.  Thus, our treatment is for a zero
temperature condensate.  It also can describe 4WM with or without the
presence of a trapping potential.

The nature of 4WM in BEC collisions of matter waves is unlike 4WM for
optical wavepacket collisions in dispersive media
~\cite{Hellwarth,Maker,Yariv}.  The nonlinearity in the case of
BEC is introduced by collisions rather than by interaction with an
external medium, and the momentum and energy constraints imposed are
different in the two cases.  The kinetic energy of massive particle
waves is quadratic in the wavevector of the particles and given by
$(\hbar{\bf k})^{2}/2m$, whereas the energy of a photon is linear in
the vacuum wavevector of the photon, ${\bf k}$, and is given by
$\hbar c|{\bf k}|$.  Moreover, the momentum of massive particle waves
is linear in the wavevector of the particles and given by $\hbar{\bf
k}$, whereas for light in a dispersive medium, it is proportional to
the product of the frequency of the light, $\omega = c|{\bf k}|$ and
the refractive index, $n(\omega)$, where the refractive index depends
upon frequency (and the propagation direction in non-isotropic media). 
Hence, conservation of energy does not in general guarantee
conservation of momentum in optical 4WM. Clearly, complications
involving the properties of an additional medium does not arise in the
BEC case.  In any case, the creation of new BEC wavepackets in 4WM is
limited to cases when momentum, energy and particle number
conservation are simultaneously satisfied.

In this paper we develop a general three-dimensional (3D) description
of four-wave mixing in single-spin-component Bose-Einstein condensates
using a mean-field approach similar to the time-dependent GPE, also
known as the nonlinear Schr\"{o}dinger equation \cite{reviews}.  We
introduce the slowly-varying-envelope approximation (SVEA), a very
powerful tool that not only gives insight into the nature of 4WM but
also gives a set of four coupled equations for the four interacting
BEC waves that are more computationally tractable for numerical
simulations of the time-dependent dynamics.  Section \ref{theory}
explains the experimental situation we have in mind and develops the
basic theoretical methods.  Section \ref{NS} describes the results of
our numerical calculations and compares these to the NIST experiment
\cite{Deng}.  Finally, in Sec.~\ref{conclusions} we present a summary
and conclusion.

\section{Theory of Matter-Wave Four-Wave Mixing}\label{theory}

In this section we describe the theoretical tools used in our study of
4WM of matter waves.  Section \ref{SecBragg} reviews how high momentum
components of a BEC can be formed using optical Bragg pulses to
prepare the initial configuration for the ``half collision'' event. 
Section \ref{SecScales} specifies the parameters that describe the
strength of the various physical effects that play a role in 4WM:
diffraction, potential energy, nonlinear self-energy, and
collisions between the different momentum wavepackets.  This Section
also describes how to transform between 1D, 2D and 3D calculations
involving the GPE. This is important because, without the
slowly-varying-envelope approximation (SVEA) that we introduce below,
full 3D calculations are too computationally expensive to carry out
for the actual experimental conditions.  Hence, the SVEA must be
explicitly checked in 2D against the full GP solution.  Section
\ref{SecSVEA} describes the details of the SVEA approximation for 4WM.
Then Section \ref{simple} introduces a simple estimate for the 4WM
output.  Finally, Section \ref{el_scat} shows how the effect of
elastic scattering between atoms in different momentum wavepackets can
be accounted for.  This process causes loss of atoms from the
wavepackets and lowers the 4WM output.

Let us consider three BEC wavepackets moving with central momenta
${\bf P}_1$, ${\bf P}_2$, and ${\bf P}_3$.  Such moving wavepackets
can be created, for example, by optically-induced Bragg diffraction of
a condensate \cite{Kozuma99}.  If these three wavepackets overlap
spatially, the self-energy of the atoms can produce matter-wave
4WM, just as the third-order Kerr type nonlinearity can produce
optical 4WM in nonlinear media.  One can imagine a number of
scenarios in which 4WM can occur in matter-wave interactions.  One
can consider a ``whole collision'' in which three initially
separated BEC wavepackets collide together at the same time, or a
``half collision'' in which the wavepackets are initially formed in
the same condensate at (nearly) the same time.  Although we considered
the ``whole collision'' case in Ref.~\cite{Tripp98}, the ``half
collision'' case is easier to realize experimentally \cite{Deng} using
the above-mentioned Bragg diffraction technique~\cite{Kozuma99}.  In
what follows, we consider only this configuration, in which the three
wavepackets initially overlap because they have been created as copies
of the initial condensate.  These wavepackets have different
non-vanishing central momenta and therefore they fly apart from one
another after they have been created.

\begin{figure} 
\centerline{\epsfxsize=4.25in\epsfbox{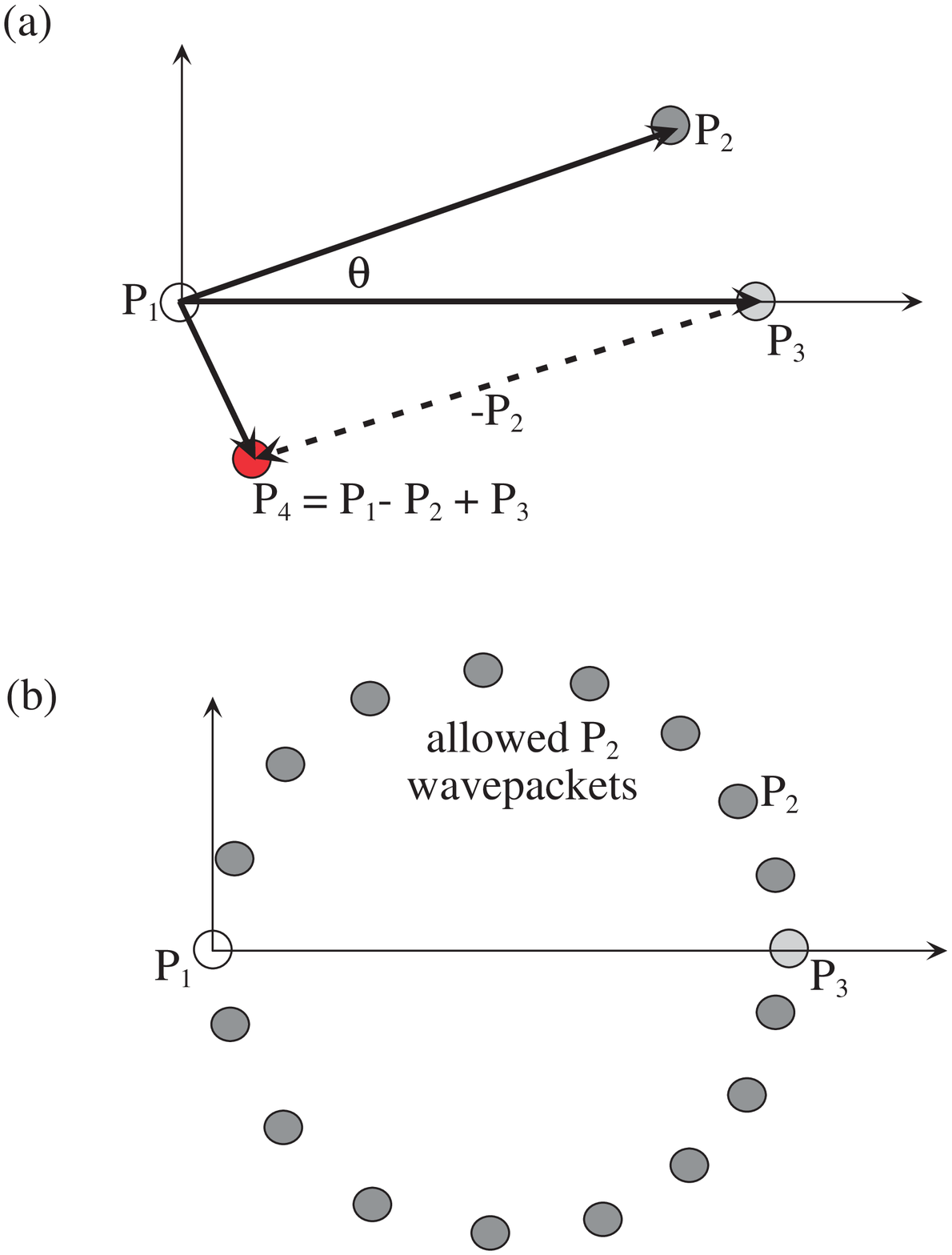}}
\caption {Momentum space view of the wavepackets participating in the
four-wave mixing process.  (a) Conservation of momentum in the
laboratory frame.  (b) A set of possible wavepackets in the laboratory
frame with momenta that satisfy the phase-matching conditions in
Section \ref{SecSVEA}, namely, $|{\bf P}_2|=|{\bf P}_3| \cos{\theta}$.}
\label{f1}
\end{figure}

\begin{figure}
\centerline{\epsfxsize=4.25in\epsfbox{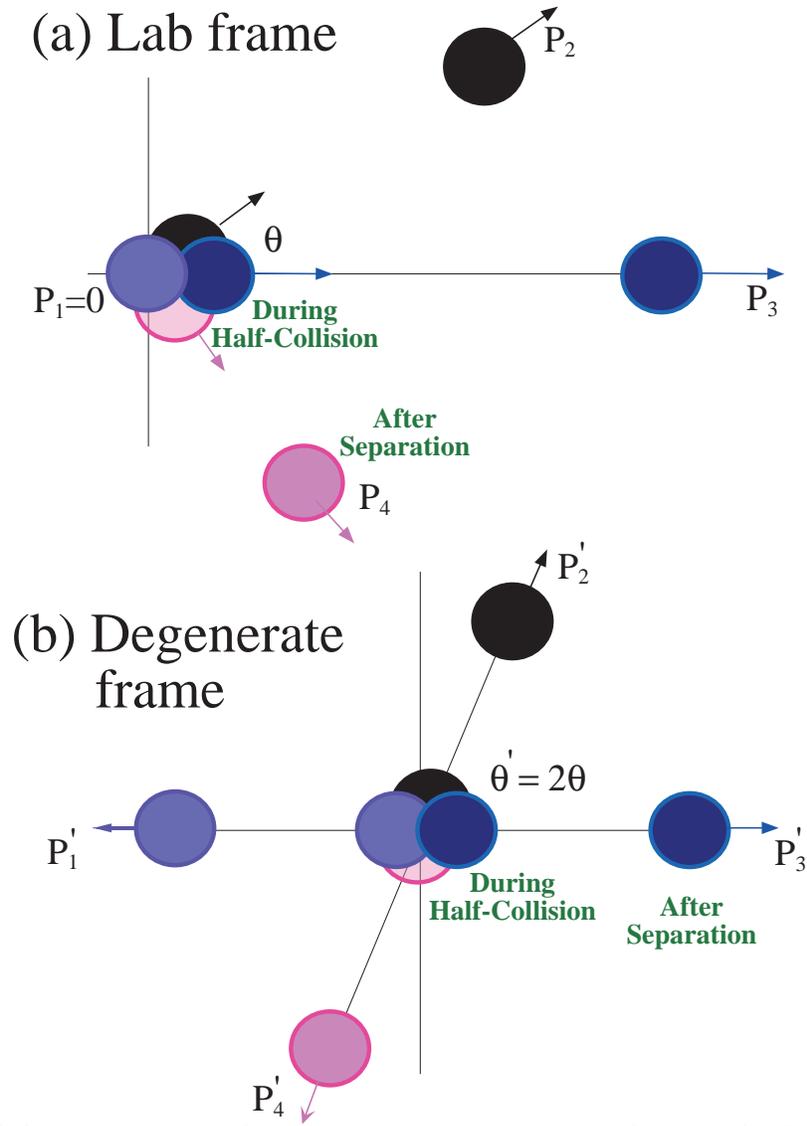}}
\caption {(a) Lab frame view of the four-wave mixing process, showing
the four wavepackets at early time while they are still interacting
and at late time after they have separated.  (b) Degenerate frame view
of the same cases as in (a).}
\label{f2}
\end{figure}

Fig.~\ref{f1}a shows the basic configuration in momentum space of the
wavepackets which we consider here.  Two daughter condensate
wavepackets with momenta ${\bf P}_2$ and ${\bf P}_3$ are created from
a parent condensate with mean momentum ${\bf P}_1=0$.  Fig.~\ref{f2}a
shows these three momenta in the lab frame in which the experiment is
carried out at two different times: during the early stage of the
``half collision'' when they still overlap spatially, and at a later
time when they have spatially separated into four distinct
wavepackets.  We let ${\bf P}_3$ lie along the $x$-axis of the
coordinate system, and ${\bf P}_2$ make some angle $\theta$ with
respect to the $x$-axis.  Nonlinear 4WM creates a fourth wavepacket
with momentum ${\bf P}_4={\bf P}_1 - {\bf P}_2 + {\bf P}_3$.  We
demonstrate below in Sec.~\ref{SecSVEA} that four-wave mixing of
matter waves is only possible if there exists a coordinate frame in
which the mixing is degenerate, that is, all four ${\bf P}'_i$ values
in this frame have the same magnitude.  Fig.~\ref{f2}b shows the
degenerate frame corresponding to a moving frame with velocity ${\bf
V}_{deg} = ({\bf P}_1 + {\bf P}_3)/(2m)$, where $m$ is the atomic
mass.  The total momentum is zero in the degenerate frame, and the
wavepackets move in oppositely moving pairs.  The angle
$\theta^{\prime}$ between the vectors ${\bf P}_2^{\prime}$ and ${\bf
P}_3^{\prime}$ is arbitrary.  In the laboratory frame, the angle
$\theta$ is given by $\theta = \theta^{\prime}/2$, and the length of
the vector ${\bf P}_2$ is given by $|{\bf P}_2| = |{\bf
P}_3|\cos(\theta)$.  Fig.~\ref{f1}b shows a set of different possible
values of ${\bf P}_2$.

\subsection{Bragg Pulse Creation of High Momentum Components}
\label{SecBragg}

We assume that the condensate has only a single spin-component, and
that its dynamics can be described by the GPE, which is known to 
provide an excellent account of condensate properties \cite{reviews}:
\begin{equation}
    i\hbar \frac{\partial\Psi({\bf r},t)}{\partial t} = (T_{{\bf r}}+
    V({\bf r},t) + NU_0|\Psi|^2) \Psi({\bf r},t), \label{GP}
\end{equation}
where $T_{{\bf r}} = \frac{-\hbar^2}{2m} \nabla_{\bf r}^2$ is the
kinetic energy operator, $V({\bf r},t)$ is the external potential
imposed on the atoms, $NU_{0} = N\frac{4\pi a_{0}\hbar^{2}}{m}$ is the
atom-atom interaction strength that is proportional to the $s$-wave
scattering length $a_{0}$ (assumed to be positive), $m$ is the atomic
mass, and $N$ is the total number of atoms.  The numerical methods for
solving the GPE are described below in Sec.~\ref{NS}.

First, we use the GPE to obtain the ground state condensate in the
trapping potential at time $t=0$, $\Psi({\bf r},t=0)$.  This
condensate wavefunction is centered around ${\bf r}={\bf 0}$, and
normalized to unity.  We assume, as is the case in the NIST
experiments~\cite{Deng}, that the trapping potential $V({\bf r},t)$ is
turned off at $t=0$ and that the condensate is allowed to evolve under
the influence of only the mean-field interaction until time $t_1$. 
This includes the special case $t_1=0$.  We could equally well treat
the case of leaving the trap on, and we would obtain similar results. 
Eq.~(\ref{GP}) determines the evolved condensate wavefunction,
$\Psi({\bf r},t_1)$.  After this period of free evolution, the Bragg
pulses are applied to create the wavepackets with momenta ${\bf
P}_{1}$, ${\bf P}_{2}$ and ${\bf P}_{3}$.  The momentum differences
$|{\bf P}_{i} - {\bf P}_{j}|$ are much larger than the momentum spread
of the initial parent BEC wavepacket.  The experimental time scale
$\delta t$ for creating these wavepackets is short ($\approx$ 70
$\mu$s) compared to the time scale on which the wavepackets evolve.
The state at $t_2=t_1+\delta t$ provides the initial condition for
subsequent evolution of these three wavepackets as they undergo
nonlinear evolution.

The initial state at $t_2$ immediately after the Bragg pulse sequences
can be approximated in a number of ways.  In principle one could set
up a set of coupled GPEs for the ground and excited atomic state
components and explicitly include the effect of coupling the light
field to the excited electronic state.  A simpler approach would be to
carry out an adiabatic elimination of the excited state and develop an
effective light-shift potential in which the ground state atoms move. 
If such approaches are carried out in this case, they show that the
light acts as a ``sudden'' perturbation such that each of the
wavepackets with central momenta ${\bf P}_{1}$, ${\bf P}_{2}$ and
${\bf P}_{3}$ is to a very good approximation simply a ``copy'' of the
parent condensate at $t=t_1$ \cite{Marya}.  Thus, the initial
condition immediately after the application of the Bragg pulses can be
approximated as being comprised of three BEC wavepackets,
\begin{equation} 
    \Psi({\bf r},t_2) = \Psi({\bf r},t_1) \sum_{i=1}^{3} f_{i}^{1/2}
    \exp(i{\bf P}_{i}\cdot{\bf r}/\hbar), \label{in_con}
\end{equation} 
where $f_i=N_i/N$ is the fraction of atoms in wavepacket $i$, and
$\sum_{i=1}^{3} f_{i} = 1$ so the norm of $\Psi$ remains unity.

After the formation of the wavepackets with momenta ${\bf P}_{1}$,
${\bf P}_{2}$ and ${\bf P}_{3}$, the initial wavefunction in
Eq.~(\ref{in_con}) evolves, and the wavepackets with the different
momenta separate.  During this separation,the nonlinear term in the
GPE generates a wavepacket with central momentum ${\bf P}_{4} = {\bf
P}_{1} - {\bf P}_{2} + {\bf P}_{3}$, as long as the constraints
discussed in relation to Figs.~\ref{f1} and \ref{f2} are
satisfied.  Energy and momentum are conserved during the wavepacket
evolution.  This can be readily checked by verifying that $dE(t)/dt =
0$ and $d{\bf P}(t)/dt = 0$, where
\begin{equation} 
    E(t) = \langle \Psi(t) |(T_{{\bf r}}+ \frac{1}{2} U_0|\Psi|^2)
    |\Psi(t) \rangle, \label{energy} 
\end{equation} 
is the energy per particle and
\begin{equation} 
    {\bf P}(t) = -i\hbar \langle \Psi(t)|\mbox{\boldmath $\nabla$}|\Psi(t) 
    \rangle \ ,
\label{momentum}
\end{equation} 
is the momentum per particle.  We have verified numerically that 
energy and momentum are indeed conserved in our calculations described 
in Section \ref{NS}.

\subsection{Characteristic Time Scales, and Dimensionless 
Parameters}\label{SecScales}

In this subsection we discuss characteristic time scales that can be
used to estimate the importance of the various effects occurring
during the dynamics for a particular set of experimental parameters. 
It is convenient to use the Thomas--Fermi (TF) 
approximation~\cite{reviews} to give
quantitative estimates of the size of the condensate and the time
scales characterizing the dynamics.  In the TF approximation, one
neglects the kinetic energy operator in the time-independent nonlinear
Schr\"{o}dinger equation,
\begin{equation} 
\mu \Psi = (T_{{\bf r}}+ V({\bf r},t) + NU_0|\Psi|^2) \Psi ,
\label{TIGP}
\end{equation}
where $\mu$ is the chemical potential, to obtain the following
analytical expression for the wavefunction: $|\Psi({\bf r})|^2 =
\frac{\mu-V({\bf r})}{NU_0}$ for ${\bf r}$ such that $V({\bf r}) \le
\mu$ and $\Psi({\bf r})=0$ otherwise.  The TF approximation is valid
for sufficiently large numbers of atoms $N$. It is convenient to define
the geometric average of the oscillator frequencies for an asymmetric
harmonic potential as $\bar{\omega} =
(\omega_x\omega_y\omega_z)^{1/3}$.  The size of the condensate is then
given by the TF radius $r_{TF} = \sqrt{2\mu/(m\bar{\omega})}$, where
the TF approximation to the chemical potential $\mu$ is determined by
the normalization of the wavefunction to unity and is given by $\mu =
\frac{1}{2} \left( \frac{15 U_0 N}{4\pi}\right)^{2/5}
(m\bar{\omega}^2)^{3/5}$.  Hence, the TF radius $r_{TF}$ scales with
$N$ as $N^{1/5}$.  The size of the TF wavepacket in the $i=$ $x$, $y$, 
and $z$ directions is $r_{TF}(i)= (\bar{\omega}/\omega_i)r_{TF}$.

In order to estimate the importance of the various terms in the GPE,
we set $V=0$ for free wavepacket evolution and rewrite Eq.~(\ref{GP})
in terms of characteristic time scales $t_{DF}$ for diffraction, and
$t_{NL}$ for the nonlinear interaction, in the following manner
\cite{Tripp98,Trip1,Trip2}:
\begin{equation}
    \frac{\partial\Psi}{\partial t} = i\left[ \frac{r_{TF}^2}{t_{DF}} \,
    (\frac{\partial^2 }{\partial x^2} + \frac{\partial^2
    }{\partial y^2} + \frac{\partial^2 }{\partial z^2}) - 
    \frac{1}{t_{NL}} \,\frac{|\Psi|^2}{|\Psi_{m}|^{2}} \right] \Psi.
\label{GP_reduced}
\end{equation}
The diffraction time and the nonlinear interaction time are given by
$t_{DF} = 2m r_{TF}^2/\hbar$, $t_{NL}=(NU_0 |\Psi_{m}|^2
/\hbar)^{-1}$, respectively.  Here $|\Psi_{m}|^2$ is the maximum value
of $|\Psi({\bf r})|^2$, i.e., $|\Psi_{m}|^2 = |\Psi({\bf 0})|^2$;
hence in the TF approximation, $t_{NL}^{-1}=\mu/\hbar$.  The smaller
the characteristic time, the larger is the corresponding term in
the GPE. We also define the collision duration time $t_{col} =
(2r_{TF})/v$, where ${\bf v}=({\bf P}_{3}-{\bf P}_{1})/m$ is the
initial relative velocity of wavepackets 1 and 3.  Thus, $t_{col}$ is
the time it takes the wavepackets 1 and 3 to move so that they just
touch at their TF radii, and therefore no longer overlap.  The ratio
$t_{col}/t_{NL}$ gives an indication of the strength of the
nonlinearity during the collision.  The larger the ratio of
$t_{col}/t_{NL}$, the stronger the effects of the nonlinearity during
the overlap of the wavepackets.  These characteristic times stand in
the ratios $t_{DF}:t_{col}:t_{NL} = 1 : \frac{\lambda}{2 \pi r_{TF}} :
\frac{r_{TF}}{6 a_{0} N}$, where $\lambda$ is the De Broglie
wavelength associated with the wavepacket velocity $v$.  Experimental
condensates with $t_{col}/t_{NL} \gg 1$ can be readily achieved. 
Thus, the nonlinear term will have time to act while the BEC
wavepackets remain physically overlapped during a collision.  Another
relevant time scale in the dynamics is the characteristic condensate
expansion time, $t_{exp} = \bar{\omega}^{-1}$.  In the typical
experiments modeled below, $t_{DF} \gg t_{exp} > t_{col} > t_{NL}$.

In addition to time scales, there are several natural length scales
that are important: the size $r_{TF}$ of the condensate, the scale
$(\Delta k)^{-1}$ of phase variation across the parent condensate as
it expands and develops a momentum spread $\hbar \Delta k$ due to the
mean field potential, and the scale $(k')^{-1}$ of phase variation due to
the fast imparted momentum $P'=\hbar k'$, where $P'$ is the common
magnitude of the momentum for the packets in the degenerate frame
(Fig.~\ref{f1}).  These stand in the relation $(k')^{-1} \ll
(\Delta k)^{-1} \ll r_{TF}$.  The grid spacings in numerical
calculations are determined by the necessity to resolve the
wavefunction on its fastest scale of variation.  Thus, using the form
of Eq.~(\ref{in_con}) for $\Psi$ requires a grid smaller than
$(k')^{-1}$.  This requirement limits practical calculations to 2
dimensions (2D).  We will introduce an approximation in the next
section that allows three dimensional (3D) calculations by eliminating
the rapidly varying phase factors from the equations to be solved.

We find it convenient to use reduced dimensionless variables to
calculate the dynamics.  The most commonly used set of reduced
dimensionless variables in BEC problems involves using ``trap units''
\cite{reviews}.  Here however, except for determining the initial
conditions at $t=0$, the trap potential is turned off, and trap units
are not particularly relevant.  Since we do both 2D and 3D
calculations, some care is needed in developing a set of units.  The
primary requirement to simulate 3D experiments with a 2D model is that
the relations between the characteristic timescales, $t_{DF}$,
$t_{col}$ and $t_{NL}$, are as determined by experiment.  We have done
this by scaling the solution of the $d$--dimensional time-dependent
GPE by a $d$--dimensional volume so that the coefficient of the
nonlinear term depends only on the dimension and the chemical
potential $\mu$.  By scaling the condensate wavefunction as $\Psi =
\bar{\Psi}/\sqrt{r_{TF}^{d}}$, the $d$--dimensional time-dependent GPE
for a harmonic potential with frequencies $\omega_j$, $j=1 \dots d$
can be written as
\begin{equation}
    i\hbar\frac{\partial\bar{\Psi}\left({\bf r}\right)}{\partial t} =
    -\frac{\hbar^{2}}{2m}\sum_{j=1}^{d}
    \frac{\partial^{2}\bar{\Psi}}{\partial x_{j}^{2}} +
    \left(\sum_{j=1}^{d}\frac{1}{2}m\omega_{j}^{2}x_{j}^{2}\right)
    \bar{\Psi}\left({\bf r}\right) 
    + \left(\frac{\pi^{d/2}}{\Gamma(2+\frac{d}{2})}\right)\mu_{{\rm TF}}
    \left|\bar{\Psi}\left({\bf r}\right)\right|^{2}
    \bar{\Psi}\left({\bf r}\right) \ .
    \label{GPd}
\end{equation}
Here $\bar{\Psi}$ is dimensionless for any $d$, and the known
$\mu_{TF}$ for the 3D problem can be transferred to an equivalent
time-dependent GPE for a 2D calculation.  Furthermore, if we
define the reduced unit of length, $x_R$, to be $x_R = r_{TF}$, define
the unit of time, $t_R$, such that $ t_R = m x_R^2/(2\hbar)$, and use
the normalization condition: $\int |\bar{\Psi}|^2  \,
d^d {\bf r}/x_R^d = 1$, we preserve the ratios between the most
important time scales of the problem.  The nonlinear time scale,
$t_{NL}$ depends only on $\mu_{TF}$ and is independent of dimension.  The
specific relations between the 3D nonlinear coupling parameter
$U_0^{3D}$ multiplying $\left|\bar{\Psi}\left({\bf r}\right)\right|^{2}
\bar{\Psi}\left({\bf r}\right)$ in Eq.~(\ref{GPd}) and $U_0^{1D}$ and
$U_0^{2D}$, the respective self-energy parameters in 1D and 2D are:
$U_0^{1D} = \frac{5}{2\pi}U_0^{3D}$, and $U_0^{2D} =
\frac{15}{16}U_0^{3D}$.  These values for $U_0^{d}$ insure that the
chemical potential $\mu_{TF}$ (and all the time scales) are the same as in
3D.

\subsection{Slowly Varying Envelope Approximation}\label{SecSVEA}

Let us consider the case when the total wavefunction consists of four
wavepackets moving with different central momenta ${\bf P}_i= \hbar
{\bf k}_i, i=1,\ldots,4$.  We write the wavefunction as
\begin{equation}
\Psi({\bf r},t) = \sum_{i=1}^{4} \Phi_i({\bf r},t)\,\exp{[i({\bf
k}_i{\bf r} - \omega_i t)]}\,,
\label{SVEApsi}
\end{equation}
in order to separate out explicitly the fast oscillating phase factors
representing central momentum $\hbar {\bf k}_i$ and kinetic energy
$E_i = \hbar\omega_i = \hbar^2 k_i^2/2m$.  The slowly varying
envelopes $\Phi_i({\bf r},t)$ vary in time and space on much longer
scales than the phases.  The number of atoms in each wavepacket is
$N_i= N \int_V|\Phi_i({\bf r},t)|^2 d^3{\bf r}$, and
$\sum_{i=1}^{4}N_{i}=N$ is a constant.  Although the slowly varying
envelope $\Phi_4({\bf r},t=0)$ is unpopulated initially, it evolves
and becomes populated as a result of the 4WM process.  If we
substitute the expanded form of the wavefunction in
Eq.~(\ref{SVEApsi}) into the GPE, collect terms multiplying the same
phase factors, multiply by the complex conjugate of the appropriate
phase factors, and neglect all terms that are not phase matched (phase
matched terms have stationary phases, do not oscillate, and satisfy
Eqs.~(\ref{mom_cons}-\ref{energy_cons}) below), we obtain a set of
coupled equations for the slowly varying envelopes $\Phi_i({\bf
r},t)$:
\begin{eqnarray} 
\left( \frac{\partial}{\partial t} + (\hbar{\bf k}_i/m) \cdot
\mbox{\boldmath $\nabla$} +
\frac{i}{\hbar}(-\frac{\hbar^{2}}{2m}\nabla^{2} + V({\bf r},t) )
\right) \Phi_i({\bf r},t) &=& -\frac{i}{\hbar} N U_0 
\sum_{i^*jj^*} \delta ({\bf k}_i+{\bf k}_{i^*}-{\bf k}_j -{\bf
k}_{j^*}) \times \nonumber \\
&& \delta(\omega_i + \omega_{i^*} - \omega_j - \omega_{j^*})
 \times \nonumber \\
&& \Phi_{j^*}({\bf r},t) \Phi_{i^*}^{*}({\bf r},t) \Phi_{j}({\bf r},t) \ ,
\label{SVEA}
\end{eqnarray}
where the delta-functions represent Kronecker delta-functions that are
unity when the argument vanishes.  Mixing between different momentum
components can result from the nonvanishing nonlinear terms in
Eq.~(\ref{SVEA}), which satisfy the phase matching constraints
required by momentum and energy conservation:
\begin{eqnarray}
{\bf k}_i + {\bf k}_{i^*} - {\bf k}_j - {\bf k}_{j^*} &=&0,
\,\,\,\,\,\,\,\,\,\, \label{mom_cons} \\
k^2_i + k^2_{i^*} - k^2_j - k^2_{j^*} &=& 0 \ .
\label{energy_cons}
\end{eqnarray}
Each of the indices $i,i^*,j,j^*$ may take any value between 1 and 4. 
Eqs.~(\ref{mom_cons}) and (\ref{energy_cons}) are automatically
satisfied in two cases: (a) $i=i^*=j=j^*$ (all indices are equal), or
(b) $j=i \ne j^*=i^*$ (two pairs of equal indices).  The corresponding
terms describe what is called in nonlinear optics cross and self
modulation terms respectively.  The cross and self phase modulation
terms do not involve particle exchange between different momentum
components.  In the absence of the trapping potential they modify both
amplitude and phase of the wavepacket through the mean field
interaction.  Particle exchange between different momentum wavepackets
occurs only when all four indices in Eq.~(\ref{SVEA}) are different,
and conservation of momentum and energy of the atoms participating in
the exchange process occurs.  A set of {\it coupled} equations
involving wave mixing between the various momentum components is
therefore obtained.

The momentum conservation of Eq.~(\ref{mom_cons}) implies ${\bf k}_i +
{\bf k}_{i^*} = {\bf k}_j + {\bf k}_{j^*} = $ {\boldmath $\kappa$}. 
It is always possible to construct a special reference frame, which we
call the {\em degenerate frame}, where {\boldmath $\kappa$}$=0$. 
Consequently, in this frame ${\bf k}_i =- {\bf k}_{i^*}$ and ${\bf
k}_j =- {\bf k}_{j^*}$.  In addition energy conservation in
Eq.~(\ref{energy_cons}) imposes the condition $|{\bf k}_j| = |{\bf
k}_i|$ in the degenerate frame.  In this frame all four momenta are
equal in magnitude and can be divided into two pairs of opposite
vectors.  This explains the use of the conjugated pairs of symbols
$(i,i^*)$ and $(j,j^*)$ in our notation.  The total number of
particles, in all wavepackets, is a conserved quantity.  The
geometrical configuration of the wavepacket momenta in the degenerate
frame are illustrated in Fig.~\ref{f2}b.  In the figure we see two
pairs of conjugate wavepackets (1,3) and (2,4).  All four momenta are
equal in magnitude and momenta ${\bf P}_1^{\prime}$ and ${\bf
P}_3^{\prime}$ are opposite as are the momenta ${\bf P}_2^{\prime}$
and ${\bf P}_4^{\prime}$.  The angle $\theta$ depicted in the figure
is completely arbitrary.  However, $\theta \approx 0$ is not allowed,
since the wavepackets would no longer be distinguishable. 
Fig.~\ref{f1}b shows a range of possible ${\bf P}_2$ values for
wavepackets in the lab frame that satisfy the phase-matching
conditions in Eqs.~(\ref{mom_cons}) and (\ref{energy_cons}).  These 
conditions only allow $|{\bf P}_2|=|{\bf P}_3| \cos{(\theta)}$.

4WM can be viewed as a process in which one particle is annihilated in
each wavepacket belonging to an initially populated pair of
wavepackets and simultaneously one particle is created in each of two
wavepackets of another pair, one of which is initially populated and
the other (wavepacket 4) is initially unpopulated.  Hence, using
Fig.~\ref{f2}b in the moving degenerate frame, 4WM removes one atom
from each of the ``pump'' wavepackets 1 and 3, and places one atom in
the ``probe'' wavepackets 2 and one atom in the 4WM output wavepacket
4.  This picture is a consequence of the nature of the nonlinear terms
in the four SVEA equations.  It is this bosonic stimulation of
scattering that mimics the stimulated emission of photons from an
optical nonlinear medium.

The full SVEA equations for 4WM are explicitly given by:
\begin{eqnarray}
&& \left( \frac{\partial}{\partial t} 
+ (\hbar {\bf k_1}/m) \cdot\mbox{\boldmath $\nabla$} 
+ \frac{i}{\hbar}(\frac{-\hbar^{2}}{2m}\nabla^{2} + V({\bf r},t) )
\right) \Phi_1({\bf r},t) = \nonumber \\
&& -\frac{i}{\hbar} N U_0 
(|\Phi_1|^2 +2|\Phi_2|^2 + 2|\Phi_{3}|^2 + 2|\Phi_{4}|^2) \Phi_1 
- \frac{i}{\hbar} N U_0\Phi_4 \Phi_2 \Phi_3^* \ ,
\label{SVEA1}
\end{eqnarray}
\begin{eqnarray}
&& \left( \frac{\partial}{\partial t} 
+ (\hbar {\bf k_2}/m) \cdot \mbox{\boldmath $\nabla$}  
+ \frac{i}{\hbar}(\frac{-\hbar^{2}}{2m}\nabla^{2} + V({\bf r},t) ) \right)
\Phi_2({\bf r},t) = \nonumber \\
&& -\frac{i}{\hbar} N U_0  
(|\Phi_2|^2 +2|\Phi_1|^2 + 2|\Phi_{3}|^2 + 2|\Phi_{4}|^2) \Phi_2 
- \frac{i}{\hbar} N U_0\Phi_4^* \Phi_1 \Phi_3  \ ,
\label{SVEA2}
\end{eqnarray}
\begin{eqnarray}
&& \left( \frac{\partial}{\partial t} 
+ (\hbar {\bf k_3}/m) \cdot \mbox{\boldmath $\nabla$}  
+ \frac{i}{\hbar}(\frac{-\hbar^{2}}{2m}\nabla^{2} + V({\bf r},t) ) \right)
\Phi_{3}({\bf r},t) = \nonumber \\
&& -\frac{i}{\hbar} N U_0 
(|\Phi_3|^2 +2|\Phi_1|^2 + 2|\Phi_{2}|^2 + 2|\Phi_{4}|^2) \Phi_3 
- \frac{i}{\hbar} N U_0\Phi_4 \Phi_1^* \Phi_2 \ ,
\label{SVEA3}
\end{eqnarray}
\begin{eqnarray}
&& \left( \frac{\partial}{\partial t} 
+ (\hbar {\bf k_4}/m) \cdot \mbox{\boldmath $\nabla$}  
+ \frac{i}{\hbar}(\frac{-\hbar^{2}}{2m}\nabla^{2} + V({\bf r},t) ) \right)
\Phi_{4}({\bf r},t) = \nonumber \\
&& -\frac{i}{\hbar} N U_0  
(|\Phi_4|^2 +2|\Phi_1|^2 + 2|\Phi_{2}|^2 + 2|\Phi_{3}|^2) \Phi_4 
- \frac{i}{\hbar} N U_0\Phi_1 \Phi_2^* \Phi_3  \ .
\label{SVEA4}
\end{eqnarray}
The left hand side of these equations describes the motion of the
wavepackets due to their kinetic and potential energies.  The right
hand side describes the effect of the phase matched nonlinear
interaction terms.  The last term on the right hand side of each of
the SVEA equations is a source term which either creates or destroys
atoms in the wavepacket being propagated.  The other terms on the
right hand side of the equations account for the self- and cross-phase
modulation.  These phase modulation terms provide an effective
potential for each wavepacket that accelerates the atoms in it and
modifies its internal momentum distribution.

Before we propagate the SVEA equations, the initial wavefunction of
the parent condensate is determined using the time-dependent GPE.
First, the propagation is in imaginary time to obtain the initial
eigenstate in the presence of the magnetic potential.  Then, after
turning off the magnetic potential, the free evolution in the absence
of a trapping potential is calculated to provide the initial condition
in Eq.~(\ref{in_con}).  This free evolution causes a spatially varying
phase to develop across the condensate as it expands in the absence of
the trapping potential.  Given the initial condition, the SVEA
equations can be used to propagate the envelope function of each
wavepacket, using the same numerical method used to propagate the
ordinary time-dependent GPE.

\subsection{Simple Approximations and Scaling with $N$}  \label{simple}

An estimate of the number of atoms that will be transferred to the 4WM
wavepacket can be developed as follows.  To get the small signal
growth at early times, multiply both sides of the dynamical equation 
for the rate of change of $\Phi_{4}$ , where for simplicity we keep 
only the 4WM term on the right hand side of the equation,
\begin{equation}
  \frac{\partial\Phi_4}{\partial t} = -\frac{i}{\hbar} N U_0 \Phi_1
  \Phi_2^* \Phi_3 \ ,  \label{approx1}
\end{equation}
by a small time increment $\delta t$ to get the growth $\delta
\Phi_{4}$ in $\Phi_{4}$ during $\delta t$:
\begin{equation}
    \delta \Phi_{4} \approx -i (f_1f_2f_3)^{1/2} \frac{NU_0}{\hbar}
    |\Psi|^{2} \Psi \delta t \approx -i (f_1f_2f_3)^{1/2} \frac{\delta
    t}{t_{NL}} \Psi \,.
\end{equation}
Here $f_i=N_i/N$ is the initial fraction of atoms in wavepacket $i$, 
and we assume that $\Phi_i=f_i^{1/2}\Psi$ at early times, because the
three wavepackets initially satisfy this relation.  Since most of the
growth takes place in the center of the packets where $\Psi$ is the
largest, the factor $ N U_0|\Psi|^2/\hbar$ is approximated by
$1/t_{NL} = N U_0|\Psi({\bf 0})|^2/\hbar$.  Upon squaring this
equation, and integrating over all space, the total growth in the 4WM
output $\delta f_4$ is
\begin{equation}
    \delta f_4 = \frac{\delta N_4}{N} \approx f_1f_2f_3 \left
    (\frac{\delta t}{t_{NL}} \right )^2 \,. \label{EarlyTime}
\end{equation}
Thus, the 4WM signal should grow quadratically at early times.
If we take $\delta t$ to be the total interaction time $t_{col}$ 
defined in Section \ref{SecScales}, then an estimate of the total 
4wm output fraction is
\begin{equation}
    f_4 = \frac{N_4(t_{col})}{N} \approx f_1f_2f_3 \left
    (\frac{t_{col}}{t_{NL}} \right )^2 \,.
\end{equation}    
This should be an upper bound on the 4WM output, since the mutual
interaction of the packets due to the self- and cross-phase modulation
terms (the self- and cross-interaction energy terms), and their
separation from one another when $t \approx t_{col}$, will lower the
output.  Using the TF approximation, $1/t_{NL}=\mu/\hbar \sim N^{2/5}$
and $t_{col}=2r_{TF}/v \sim N^{1/5}$.  Thus, the output fraction
$\frac{N_4}{N} \sim (N^{1/5}N^{2/5})^2$ scales as $N^{6/5}$.  This
scaling, which was discussed in reference \cite{Deng}, will be checked
in our numerical calculations below.

\subsection{Elastic scattering loss}\label{el_scat}

Atoms from two {\it different} momentum wavepackets can undergo
$s$-wave elastic scattering that removes the atoms from the packets
and scatters them into $4\pi$ steradians \cite{BTBJ}.  This becomes
important when the mean-free-path $\ell_{mfp}$ becomes comparable to
or smaller than the condensate size, $r_{TF}$.  The mean-free-path is
$\ell_{mfp} = (\sigma {\bar n})^{-1}$, where $\sigma = 8\pi a_0^2$ is
the elastic scattering cross section and ${\bar n}$ is the mean
density.  Profuse elastic scattering of this type has been recently
observed \cite{SK-K}.  This mechanism can also affect the 4WM process
since loss of atoms from the moving packets reduce the nonlinear
source terms in the SVEA equations.  Although the cloud of elastically
scattered atoms can not be simply described by the mean-field picture,
the loss of atoms from the wavepackets due to this elastic scattering
mechanism can be described in terms of the SVEA. This is because each
momentum component is treated separately, and the loss terms due to
elastic scattering can be added to the SVEA equations.

The elastic scattering loss is incorporated by adding loss terms to
the right hand side of the envelope equations in the form of
imaginary potentials that are proportional to the density of the
``other'' momentum component involved in the elastic scattering.  The
full SVEA equations for 4WM, including the effects of elastic
scattering loss \cite{BTBJ}, are given by:
\begin{eqnarray}
&& \left( \frac{\partial}{\partial t}  
+ (\hbar{\bf k_1}/m) \cdot \mbox{\boldmath $\nabla$}
+ \frac{i}{\hbar}(\frac{-\hbar^{2}}{2m}\nabla^{2} + V({\bf r},t) ) \right)
\Phi_1({\bf r},t) =  \nonumber \\
&& -\frac{i}{\hbar} N U_0 
(|\Phi_1|^2 +2|\Phi_2|^2 + 2|\Phi_{3}|^2 + 2|\Phi_{4}|^2) \Phi_1 
- \frac{i}{\hbar} N U_0\Phi_4 \Phi_2 \Phi_3^* \nonumber \\
&&- \frac{(\hbar|{\bf k}_1-{\bf k}_2|/m) \sigma N}{2} |\Phi_2|^2 \Phi_1 
- \frac{(\hbar|{\bf k}_1-{\bf k}_3|/m) \sigma N}{2} |\Phi_3|^2 \Phi_1 
- \frac{(\hbar|{\bf k}_1-{\bf k}_4|/m) \sigma N}{2} |\Phi_4|^2 \Phi_1 \ ,
\label{SVEA1el}
\end{eqnarray}
\begin{eqnarray}
&& \left( \frac{\partial}{\partial t} 
+ (\hbar{\bf k_2}/m) \cdot \mbox{\boldmath $\nabla$}  
+ \frac{i}{\hbar}(\frac{-\hbar^{2}}{2m}\nabla^{2} + V({\bf r},t) ) \right)
\Phi_2({\bf r},t) = \nonumber \\
&& -\frac{i}{\hbar} N U_0  
(|\Phi_2|^2 +2|\Phi_1|^2 + 2|\Phi_{3}|^2 + 2|\Phi_{4}|^2) \Phi_2 
- \frac{i}{\hbar} N U_0\Phi_4^* \Phi_1 \Phi_3 \nonumber \\
&&- \frac{(\hbar|{\bf k}_2-{\bf k}_2|/m) \sigma N}{2} |\Phi_1|^2 \Phi_2 
- \frac{(\hbar|{\bf k}_2-{\bf k}_3|/m) \sigma N}{2} |\Phi_3|^2 \Phi_2 
- \frac{(\hbar|{\bf k}_2-{\bf k}_4|/m) \sigma N}{2} |\Phi_4|^2 \Phi_2 \ ,
\label{SVEA2el}
\end{eqnarray}
\begin{eqnarray}
&& \left( \frac{\partial}{\partial t} 
+ (\hbar{\bf k_3}/m) \cdot \mbox{\boldmath $\nabla$}  
+ \frac{i}{\hbar}(\frac{-\hbar^{2}}{2m}\nabla^{2} + V({\bf r},t) ) \right)
\Phi_{3}({\bf r},t) = \nonumber \\
&& -\frac{i}{\hbar} N U_0  
(|\Phi_3|^2 +2|\Phi_1|^2 + 2|\Phi_{2}|^2 + 2|\Phi_{4}|^2) \Phi_3 
- \frac{i}{\hbar} N U_0\Phi_4 \Phi_1^* \Phi_2 \nonumber \\
&&- \frac{(\hbar|{\bf k}_3-{\bf k}_1|/m) \sigma N}{2} |\Phi_1|^2 \Phi_3 
- \frac{(\hbar|{\bf k}_3-{\bf k}_2|/m) \sigma N}{2} |\Phi_2|^2 \Phi_3 
- \frac{(\hbar|{\bf k}_3-{\bf k}_4|/m) \sigma N}{2} |\Phi_4|^2 \Phi_3 \ ,
\label{SVEA3el}
\end{eqnarray}
\begin{eqnarray}
&& \left( \frac{\partial}{\partial t} 
+ (\hbar{\bf k_4}/m) \cdot \mbox{\boldmath $\nabla$}  
+ \frac{i}{\hbar}(\frac{-\hbar^{2}}{2m}\nabla^{2} + V({\bf r},t) ) \right)
\Phi_{4}({\bf r},t) = \nonumber \\
&& -\frac{i}{\hbar} N U_0 
(|\Phi_4|^2 +2|\Phi_1|^2 + 2|\Phi_{2}|^2 + 2|\Phi_{3}|^2) \Phi_4 
- \frac{i}{\hbar} N U_0\Phi_1 \Phi_2^* \Phi_3  \nonumber \\
&&- \frac{(\hbar|{\bf k}_4-{\bf k}_1|/m) \sigma N}{2} |\Phi_1|^2 \Phi_4 
- \frac{(\hbar|{\bf k}_4-{\bf k}_2|/m) \sigma N}{2} |\Phi_2|^2 \Phi_4 
- \frac{(\hbar|{\bf k}_4-{\bf k}_3|/m) \sigma N}{2} |\Phi_3|^2 \Phi_4 \ .
\label{SVEA4el}
\end{eqnarray}
There are three elastic scattering loss terms for each SVE momentum
component $\Phi_i$ arising from the interaction of each momentum
component with the other three momentum components.  The factor of
$\frac{1}{2}$ in the loss terms is due to the fact that these are
equations for the amplitudes, not the densities.

The density dependence of the elastic scattering loss terms is
identical to that of the mean-field interaction terms since both terms
are due to elastic scattering.  It is of interest to compare the
strength (size of the coefficient) of the loss term due to elastic
scattering with the nonlinear term in the GPE. The nonlinear term has
a coefficient $U_0 / \hbar = 4\pi \hbar a_0/m$, whereas the loss term
for interaction of packets $i$ and $j$ has a coefficient $\frac{1}{2}
v \sigma = 4\pi \hbar |{\bf k}_i-{\bf k}_j| a_0^2/m$, where $v$ is the
relative velocity.  The ratio ${\cal R}= (\frac{1}{2} v \sigma)/(U_0 /
\hbar)$ of loss to mean-field terms for packets 1 and 3 in
Fig.~\ref{f1} is
\begin{equation}
    {\cal R} = 2|{\bf k}_1| a_0  \ .    
    \label{ratio}
\end{equation} 
This ratio is about $0.06$ for the NIST 4WM experiment \cite{Deng}.

\section{Numerical Simulations} \label{NS}

\subsection{Experimental Configuration}

In the NIST experiment \cite{Deng}, the initial sodium $F, M_{F} = 1,
-1$ condensate is comprised of magnetically confined atoms in a TOP
(time-orbiting-potential) trap without a discernible non-condensed
fraction.  The trap is adiabatically expanded to reduce the trap
frequencies in the $x$, $y$ and $z$ directions to 84, 59 and 42 Hz (the
frequency ratios are $\omega_x:\omega_y:\omega_z = 1:1/\sqrt{2}:1/2$). 
After adiabatic expansion, the trap is switched off by removing the
confining magnetic fields.  The condensate freely expands during a
delay time $t_1=600$ $\mu$s, after which a sequence of two Bragg
pulses of $589$ nm wavelength creates the two moving wavepackets 2 and
3.  Each 30 $\mu$s Bragg pulse is composed of two linearly polarized
laser beams detuned from the $3S_{1/2}, F = 1, M_{F} = -1
\,\rightarrow \, 3P_{3/2}, F=2, M_{F} = 2$ transition by about
$\Delta/2\pi = -2$ GHz to suppress spontaneous emission and scattering
of the optical waves by the atoms.  The frequency difference between
the two laser beams of a single Bragg pulse is chosen to fulfill a
first-order Bragg diffraction condition that changes the momentum
state of the atoms without changing their internal state.  The first
Bragg pulse is composed of two mutually perpendicular laser beams of
frequencies $\nu_{\alpha}$ and $\nu_{\beta}= \nu_{\alpha}- 50$ kHz,
and wavevectors and ${\bf k}_{\alpha} = k \hat{{\bf x}}$ and
$k_{\beta} = k \hat{{\bf y}}$.  This pulse sequence causes a fraction
$f_2$ of the BEC atoms to acquire momentum ${\bf P}_{2} = \hbar({\bf
k}_{\alpha}-{\bf k}_{\beta}) = \hbar k (\hat{{\bf x}} + \hat{{\bf
y}})$.  A second set of Bragg pulses is applied 20 ms after the end of
the first Bragg pulse sequence.  This pulse is composed of two
counter-propagating laser beams with frequencies $\nu_{\alpha}$ and
$\nu_{\beta}= \nu_{\alpha}- 100$ kHz, and wavevectors and ${\bf
k}_{\alpha} = k \hat{{\bf x}}$ and ${\bf k}_{\beta} = -k \hat{{\bf
x}}$.  This pulse sequence causes a fraction $f_3$ of the BEC atoms to
acquire momentum ${\bf P}_{3} = \hbar({\bf k}_{\alpha}-{\bf
k}_{\beta}) = 2\hbar k \hat{{\bf x}}$.  Thus, there are three initial
condensate wavepackets with momenta ${\bf P}_{1} = {\bf 0}$, ${\bf
P}_{2}$ and ${\bf P}_{3}$ as shown in Fig.~\ref{f1}.  The respective
wavepacket populations, $f_1=1-f_2-f_3$, $f_2$, and $f_3$, have a
typical ratio $f_1:f_2:f_3 = 7:3:7$.

The number of atoms could be varied between around $3\times10^5$ and
$3\times 10^6$.  As a typical example, we take $N=1.5\times10^6$ atoms
in the trap.  Taking $a_0=2.8$ nm \cite{Tiesinga96}, the nonlinear
time is $t_{NL} = 96.2$ $\mu$s.  The Thomas Fermi radius is $r_{TF} =
20.3$ $\mu$m.  Since the separation velocity defined in Section
\ref{SecScales} is $v=0.0691$ m/s for light of wavelength $589$ nm,
the physical separation time $t_{col} =\frac{2 r_{TF}}{v} = 687$
$\mu$s in the NIST experiment, and indeed is longer than the nonlinear
time.  The characteristic condensate expansion time, $t_{exp} =
\bar{\omega}^{-1} = 1.89$ ms for a trap with
$\bar{\omega}=2\pi\frac{84}{\sqrt{2}}$ s$^{-1}$.  The characteristic
diffraction time $t_{DF} = 2m r_{TF}^2/\hbar = 300$ ms provides by far
the longest time scale in the dynamics.  Thus, there is negligible
diffraction on the time scale of the experiment.

\subsection{Simulations of the NIST Experiments}\label{results}

\begin{figure}
\centerline{\epsfxsize=4.25in\epsfbox{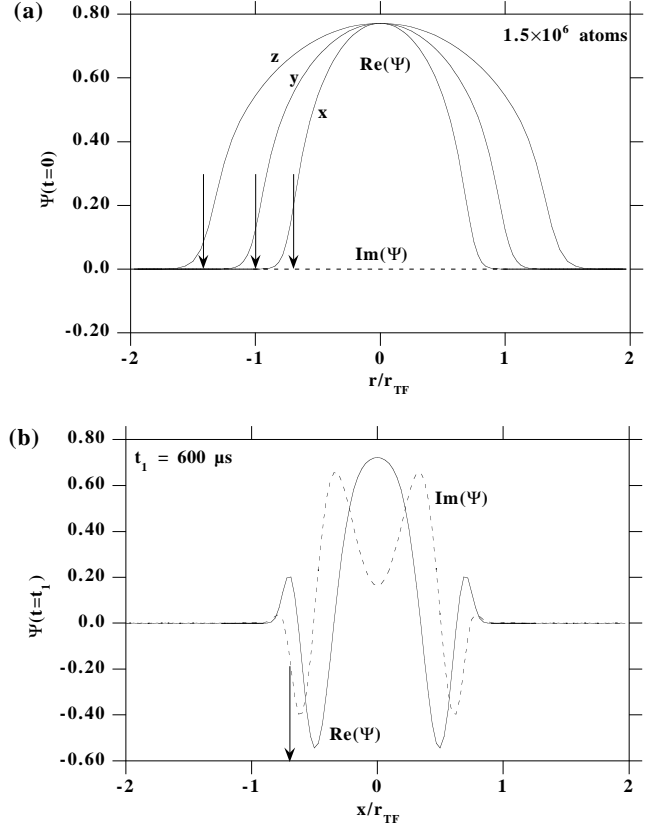}}
\caption {(a) Cuts along the $x$, $y$ and $z$ axes of the parent
condensate wavefunction $\Psi(x,y,z,t=0)$ for $N=1.5\times 10^6$ atoms
in a trap with harmonic frequencies of 84 Hz, 59.4 Hz, and 42 Hz in
the respective $x$, $y$, and $z$ directions.  The arrows show the TF
radii $r_{TF}(i)$ in the $i=x,y,z$ directions.  The curves labeled
``$x$'', ``$y$'', and ``$z$'' respectively represent
$\mathrm{Re}[\Psi(x,0,0,0)]$, $\mathrm{Re}[\Psi(0,y,0,0)]$, and
$\mathrm{Re}[\Psi(0,0,z,0)]$; $\mathrm{Im}[\Psi(x,y,z,0)]$ is
identically zero for each case.  (b) Cuts along the $x$ axis of
$\mathrm{Re}[\Psi(x,0,0,t = t_1)]$ and $\mathrm{Im}[\Psi(x,0,0,t =
t_1)]$ for $t_1 = 600\ \mu$s.}
\label{f3}
\end{figure}

\begin{figure}
\centerline{\epsfxsize=4.25in\epsfbox{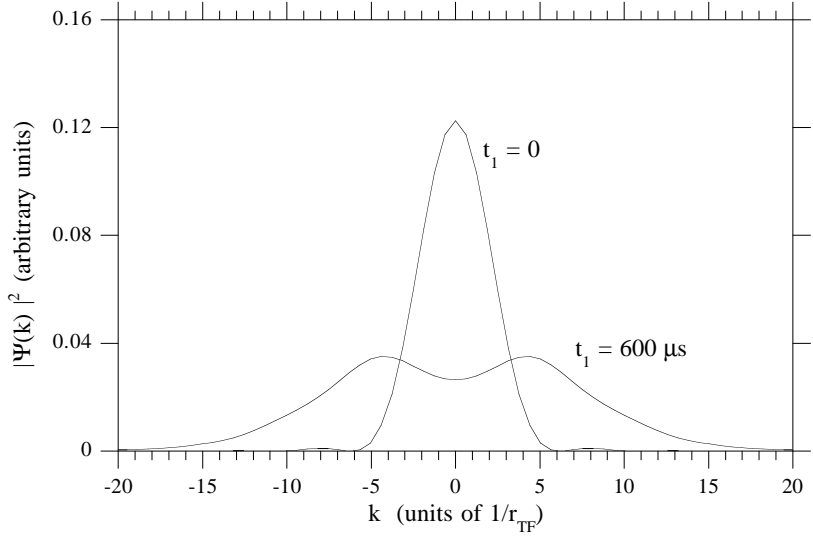}}
\caption {Cut in the $k_x$ direction $(k_y=k_z=0)$ of the squared
momentum distribution $|\Psi({\bf k},t)|^2$ for the wavefunctions in
Fig.~\ref{f3} for $t_1=0$ and $t_1=600$ $\mu$s. }
\label{f4}
\end{figure}

Our solution to the time-dependent GPE uses a standard split-operator
fast Fourier transform method to propagate an initial state forward in
time\cite{SplitFFT}.  The initial state $\Psi({\bf r},t=0)$ of the
condensate in the trap is found by iteratively propagating in
imaginary time.  Fig.~\ref{f3} shows examples of a 3D
parent condensate wavefunction $\Psi(x,y,z,t)$ for two different
times.  The $t=0$ solution shows the wavefunction in the harmonic
trap, and the $t = t_1=600$ $\mu$s solution shows the wavefunction
after 600 $\mu$s of free evolution without a trap potential.  Although
the $t=0$ wavefunction in Fig.~\ref{f3}a has a constant phase (taken
to be 0), it is apparent from Fig.~\ref{f3}b that the evolution leads
to the development of phase modulation across the condensate, i.~e.,
the wavefunction develops a spatially dependent phase, and therefore
an imaginary part of the wavefunction.  This is due to the evolution
of the condensate under the influence of the mean field term, $N U_0
|\Psi({\bf r},t)|^2$, when the trapping potential is no longer
present.  An analytic form for the spatially dependent phase which
evolves can be obtained in the Castin-Dum model \cite{CD}.  As we show
below, this phase modulation is important for 4WM. There is very little
physical expansion of the condensate after 600 $\mu$s, since the
condensate densities $|\Psi({\bf r},t)|^2$ are nearly the same for the
wavefunctions in Figs.~\ref{f3}a and \ref{f3}b.  However,
Fig.~\ref{f4} shows that the acceleration due to the mean field is
already quite evident in the momentum distribution at $t=600$ $\mu$s,
which is much broader than that at $t=0$.  The two peaks near $k=\pm 5
r_{TF}^{-1}$ in the $t=t_1=600$ $\mu$s distribution indicate the
formation of accelerated condensate particles which will lead to
condensate expansion at later times.

Our treatment for applying Bragg pulses uses the model given by
Eq.~(\ref{in_con}).  This approximation neglects detailed dynamics
during the application of the Bragg pulses.  Each initial wavepacket
$i$ at time $t_2$ after the Bragg pulses is a copy of the parent
condensate wavefunction at $t=t_1$ with population fraction
$f_i=N_i/N$.  Unless stated otherwise, we will always use the ratio
$f_1:f_2:f_3 = 7:3:7$ of population fractions as typical of the NIST
experiment \cite{Deng}.  We let the three BEC wavepackets evolve for
$t > t_2 \approx t_1$ using three different versions of the
time-dependent GPE. Two of them are 2D versions, and one is the
3D-SVEA version.  The 2D-full version uses the GPE, Eq.~(\ref{GP}), to
evolve the initial state $\Psi$ in Eq.~(\ref{in_con}).  The 2D-SVEA
version uses the SVEA form in Eqs.~(\ref{SVEA1})-(\ref{SVEA4}) for the
evolution.  A typical 2D calculation used a grid of discrete $x,y$
points within a box $5r_{TF}$ wide in the $x$ and $y$ directions
centered on $x=y=0$.  In order to resolve the rapid phase variations
due to the $e^{i({\bf k} \cdot {\bf r})}$ factor, the 2D-full
calculation required an $x,y$ grid of up to $4096\times 4096$ points. 
On the other hand, the 2D-SVEA only requires a $128\times 128$ $x,y$
grid to achieve comparable accuracy.  The 3D-SVEA calculations added a
$4r_{TF}$ wide box in the $z$ direction, and an $x,y,z$ grid of
$128\times 128\times 64$ was sufficient.

\begin{figure}
\centerline{\epsfxsize=4.25in\epsfbox{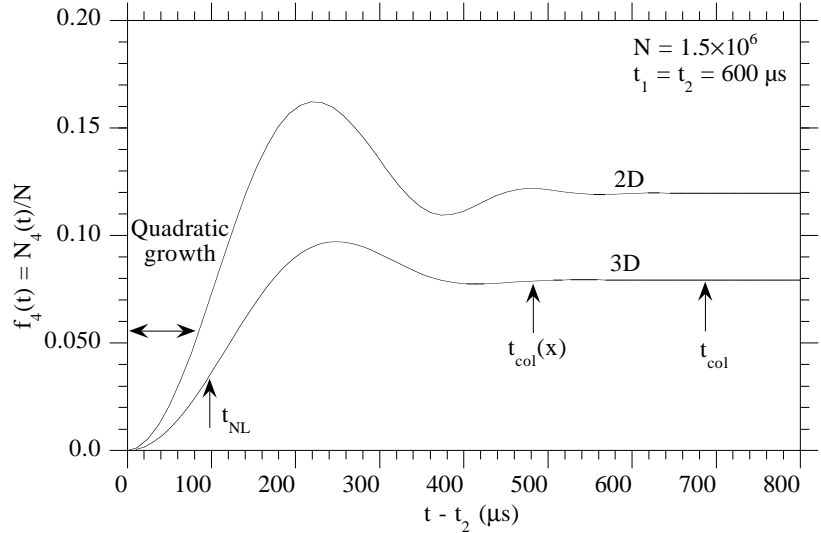}}
\caption {Comparison of $N_{4}(t)/N$ versus $t-t_2$ for 2D and 3D
calculations for $1.5\times 10^6$ atoms.  The trap is the same as in
Fig.~\ref{f3}.  The Bragg pulses are applied 600 $\mu$s after the
trapping potential is turned off and are over at time $t_2$.}
\label{f5}
\end{figure}

Fig.~\ref{f5} compares the 4WM output fraction $f_4(t)\equiv N_4(t)/N$
for the three different types of calculation for the case of
$N=1.5\times 10^6$ atoms.  The 2D-full and 2D-SVEA calculations give
the same results within numerical accuracy and can not be
distinguished on the graph.  We take this to be a strong justification
of the SVEA, and a strong indication that it will be equally
trustworthy in the 3D calculations.  In both 2D and 3D cases, the
output grows quadratically at early time, as predicted by
Eq.~(\ref{EarlyTime}).  The arrows indicate the characteristic
nonlinear time $t_{NL}$ and the collision time $t_{col}$.  In
addition, the figure shows $t_{col}(x)=t_{col}/\sqrt{2}$.  The latter
is the time it takes wavepackets 1 and 2 to move so that they just
touch at their Thomas-Fermi radii in the $x$ direction.  At that time
wavepackets 1 and 2 no longer have significant overlap with each
other, although they still have some overlap with wavepacket 3.  As
the wavepackets begin to move apart, the output saturates near $t-t_2
\approx t_{col}(x)/2$ and approaches its final value when $t - t_{2}
\approx t_{col}$.  There is a significant difference between the
3D-SVEA and 2D-SVEA output fraction.  The 4WM output is lower for the
3D case.  This is because the nonlinear 4WM process depends on the
spatial overlap of the moving wavepackets.  The packets are not as
well-overlapped geometrically in 3D as in the 2D model.  Henceforth,
all our calculations are 3D-SVEA ones, unless stated otherwise.

\begin{figure}
\centerline{\epsfxsize=4.25in\epsfbox{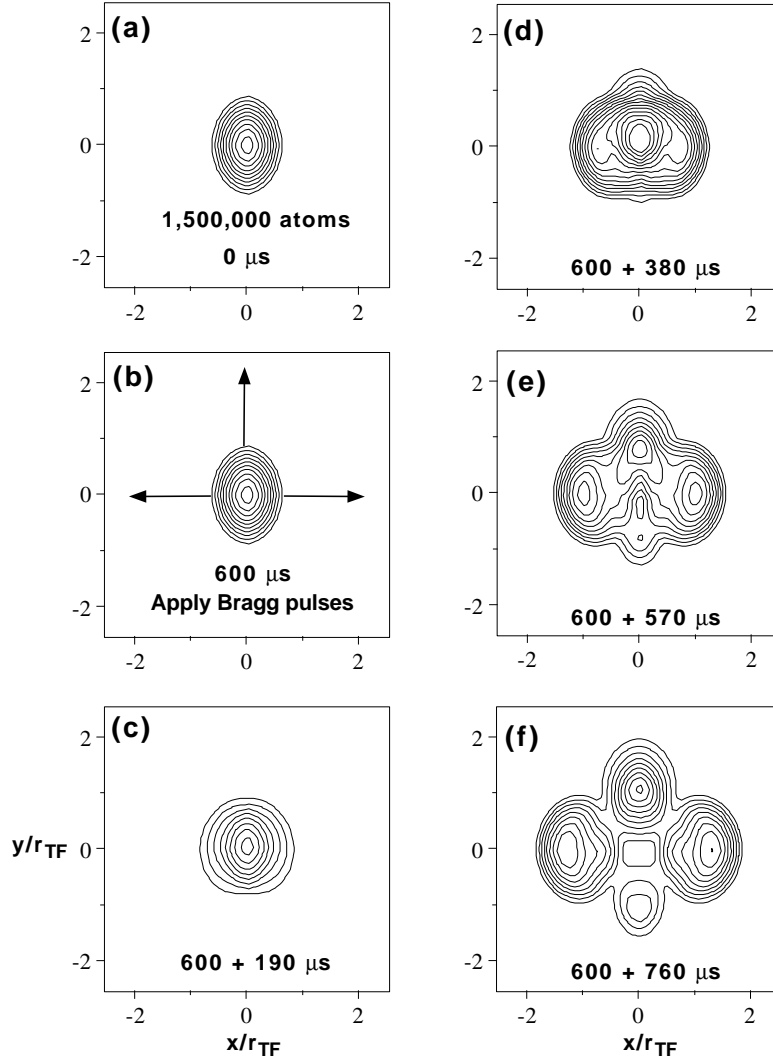}}
\caption {Contour plots of integrated column density from the 3D-SVEA
calculations vs $x$ and $y$ for $N=1.5\times 10^{6}$ and the same trap
as for Fig.~\ref{f3}.  Panels (a) through (f) show the time
development of the wavepackets from the from the time the trap is
turned off until the wavepackets physically separate.}
\label{f6}
\end{figure}

Fig.~\ref{f6} shows a sequence of contour images of the time evolution
of the wavepackets from the time the trap is turned off at $t=0$ to
the time of separation of the four wavepackets.  The contours show the
$z$-integrated column density, $\sum_{i=1}^4 \int \Phi_i(x,y,z,t)|^2
dz$, from the 3D-SVEA calculation.  (The constructive and destructive
interference fringes in the wavepacket overlap region due to the
$e^{i{\bf k} \cdot {\bf r}}$ phase factors is not shown since it would
require very high resolution to represent it with sufficient
accuracy).  Panel (a) shows the eigenstate density in the harmonic
trap.  Panel (b) shows the wavepacket at $t = t_2$ just after the
Bragg pulses have fired.  Since there is negligible expansion in the
density profile during the initial 600 $\mu$s of free evolution, the
wavepacket is very similar to that in panel (a).  However, we learned
from Fig.~\ref{f3} that a phase modulation has developed across the
wavepacket.  This does not show up in the density profile.  Panel (c)
for $t-t_2 =190$ $\mu$s indicates some initial motion by the moving
wavepackets.  In panel (d) the spread of the three wavepackets due to
their different momenta is evident, and in panel (e) the separation of
the 4WM wavepacket is clearly apparent.  Panel (e) shows the four
wavepackets after almost complete separation at $t-t_2 =760$ $\mu$s,
which is larger than $t_{col} = 687 \mu$s.

\begin{figure}
\centerline{\epsfxsize=4.25in\epsfbox{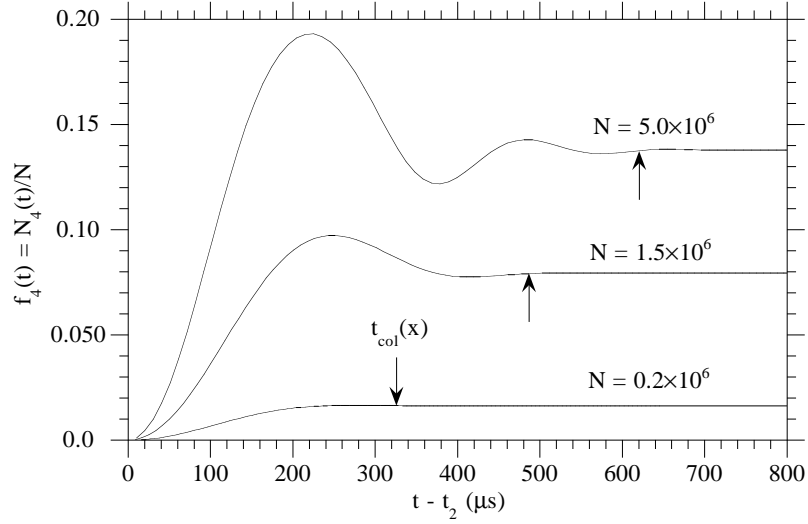}}
\caption {Comparison of $N_{4}(t)/N$ versus $t-t_2$ for $0.2\times
10^6$, $1.5\times 10^6$ and $5.0\times 10^6$ atoms.  The trap is the
same as in Fig.~\ref{f3}.  The Bragg pulses are applied 600 $\mu$s
after the trapping potential is turned off.}
\label{f7}
\end{figure}

Fig.~\ref{f7} compares the output fraction $N_4(t)/N$ versus time for
three different initial total atom numbers, $N=0.2\times 10^6$,
$1.5\times 10^{6}$ and $5.0\times 10^{6}$, and $t_1 = 600$ $\mu$s. 
Again, at early times the quadratic dependence of the fraction as a
function of time is clearly evident.  After a quadratic rise at early
time, the output saturates and even undergoes oscillations before
finally settling down to a final value when $t > t_{col}$.  The
oscillations of $N_4(t)/N$ in time develop and become more pronounced
as the initial number of atoms increases.  These are due to
back-transfer from the $i=$ 2 and 4 packets to the $i=$ 1 and 3
packets due to the mutual coupling between the packets.  A closer
examination of the detailed time evolution shows that the transfer
occurs on the trailing edge of the wavepackets where they are still
substantially overlapped.  When $N$ is large enough, the wavepackets
experience significant distortion in shape by the time they separate. 
The output fraction $N_4(t)/N$ clearly increases with $N$.

\begin{figure}
\centerline{\epsfxsize=4.25in\epsfbox{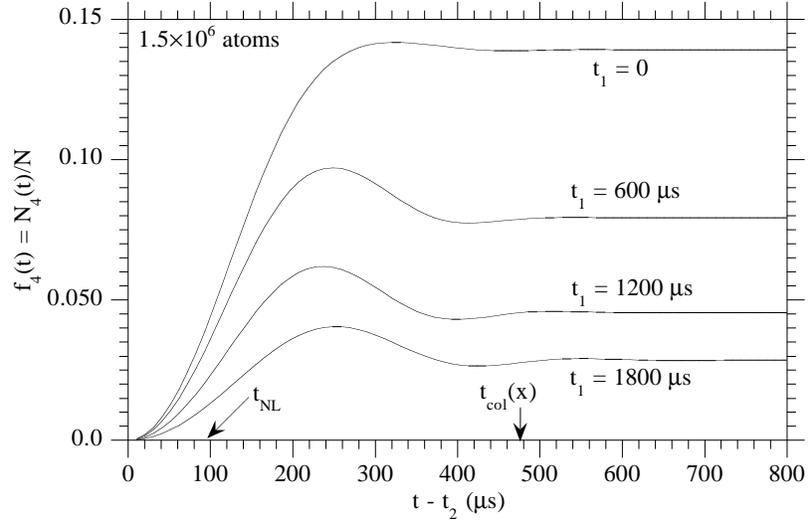}}
\caption {Comparison of $N_{4}(t)/N$ versus $t-t_2$ for $1.5\times
10^6$ atoms.  The different curves show cases where the Bragg pulses
are applied at $t_1 = 0$, $600$, $1200$ and $1800$ $\mu$s after the
trapping potential is turned off ($t_2 \approx t_1$).  The trap is the
same as in Fig.~\ref{f3}.}
\label{f8}
\end{figure}

Fig.~\ref{f8} shows the output fraction $N_4(t)/N$ versus time for
$1.5\times 10^{6}$ atoms for four different values of the free
evolution time $t_1 = 0$ $\mu$s, 600 $\mu$s, 1200 $\mu$s, and 1800
$\mu$s.  The self-phase modulation resulting from the nonlinear
self-energy interaction reduces the 4WM output as $t_1$ increases. 
This is analogous to the destruction of third harmonic generation due
to self- and cross-phase modulation in nonlinear optics \cite{Band90},
and occurs because the phase modulation destroys the phase matching
that is necessary for 4WM to develop.  For $t > t_{col}$,
the number of atoms in the different wavepackets no longer change,
since the wavepackets are well separated (exchange of the number of
bosonic atoms between wavepackets can no longer occur when the terms
in the dynamical equations responsible for 4WM vanish).  From these
calculations it seems clear that 4WM should be much stronger if the
trap is left on instead of being turned off.  These calculations
indicate that the 4WM output of the NIST experiment \cite{Deng} might
be as much as a factor of two higher if there had not been 600 $\mu$s
of free evolution before the Bragg pulses were applied.

\begin{figure}
\centerline{\epsfxsize=4.25in\epsfbox{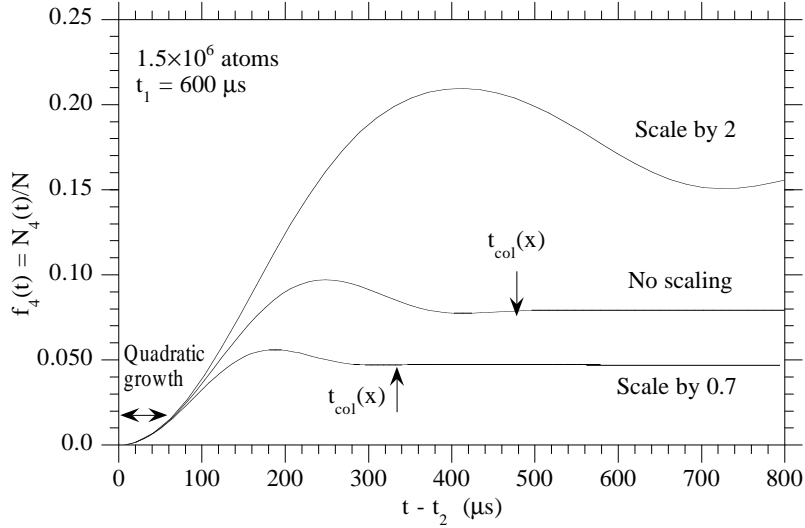}}
\caption {Comparison of $N_{4}(t)/N$ versus $t-t_2$ for $1.5\times
10^6$ atoms.  The trap is the same as in Fig.~\ref{f3}.  The Bragg
pulses are applied 600 $\mu$s after the trapping potential is turned
off.  The three different curves are for the cases where the
separation times are scaled by factors of 0.7, 1, and 2 by scaling the
separation velocities by $1/0.7$, 1, and $1/2$.}
\label{f9}
\end{figure}

We expect the 4WM output will be larger if the wavepackets stay
together for a longer interaction time $t_{col}$.  The interaction
time can be changed by changing the velocity of the wavepackets. 
Fig.~\ref{f9} plots $N_{4}(t)/N$ versus time for $1.5\times 10^6$
atoms for the original case shown in Figs.~\ref{f7} and \ref{f8} and
for two new cases where the interaction times are changed by factors
of 0.7 and 2.  This is achieved in the code by scaling the momentum
wavevectors by factors of $1/0.7$ and $1/2$ respectively.  Our
calculations show that the 4WM output is reduced by a factor of 0.6 in
the first case and increased by a factor of 2 in the second.  In
principle, velocities of the wavepackets can be controlled by changing
the frequencies and angle of the two Bragg pulses that create an
outcoupled wavepacket \cite{Kozuma99}.  Thus, some degree of control
over the 4WM output should be possible by varying the interaction
time.

\begin{figure}
\centerline{\epsfxsize=4.25in\epsfbox{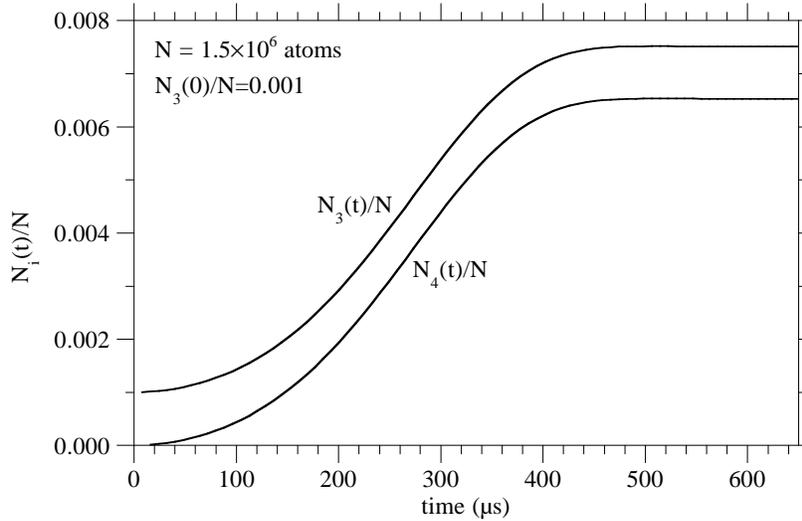}}
\caption {Growth of $N_{4}(t)/N$ and $N_{4}(t)/N$ versus $t-t_2$ for
the case where a weak probe wavepacket 2 with initial population
fraction 0.001 encounters strong ``pump'' wavepackets with initial
fractions 0.4995.  The trap is the same as in Fig.~\ref{f3}.  The
Bragg pulses are applied 600 $\mu$s after the trapping potential is
turned off.}
\label{f10}
\end{figure}

Fig.~\ref{f10} shows $f_3(t)$ and $f_4(t)$ for the case of a weak
$i=2$ ``probe'' with initial population fraction $0.001$ incident on
two strong $i=$ 1 and 3 ``pump'' wavepackets with population fractions
0.4995.  This is analogous to the phase conjugation process envisioned
in reference \cite{Goldstein95}.  Here bosonic stimulation, which
removes 2 atoms from the ``pump'' packets 1 and 3 and puts them in
packets 2 and 4, results in a strong amplification of packet 2, which
grows in atom number 8-fold as the 4WM signal grows.

\begin{figure}
\centerline{\epsfxsize=4.25in\epsfbox{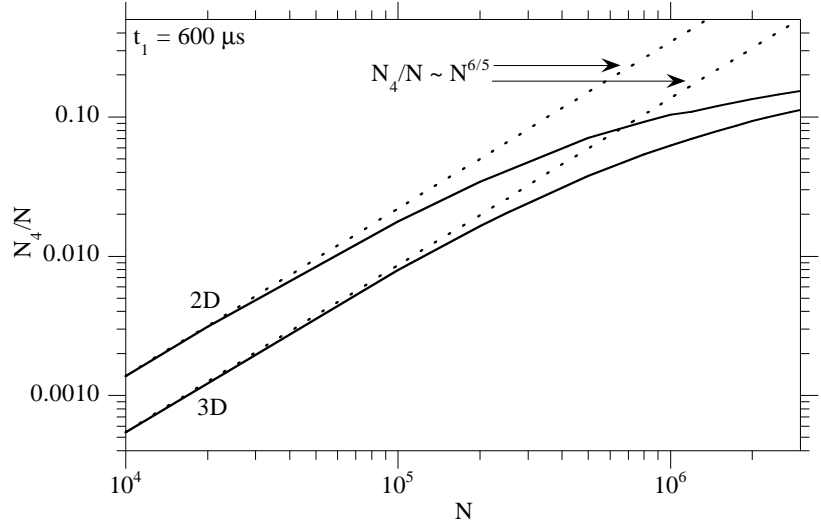}}
\caption {$N_{4}/N$ dependence dependence on the total number of
atoms, $N$, calculated in 2D and 3D. The dashed lines show the
$N^{6/5}$ dependence predicted by the simple theory in subsection
\ref{simple}.  The trap is the same as in Fig.~\ref{f3}.  The Bragg
pulses are applied 600 $\mu$s after the trapping potential is turned
off.}
\label{f11}
\end{figure}

Fig.~\ref{f11} shows 4WM output fraction $N_4/N$ after the
half-collision is over ($t>t_{col}$) as a function of $N$, plotted in
a log--log plot.  The figure shows the results for both the 2D-SVEA
and 3D-SVEA calculations.  The dashed lines show the 4WM output for
small $N$ scales well with $N^{6/5}$, as estimated from the simple
model in Section \ref{simple}.  The scaling with $N^{6/5}$ for small
$N$ is clearly evident in both 2D and 3D results.  The latter is
uniformly lower than the former, due to the smaller overlap of the
wavepackets in 3D because of geometrical reasons, but saturates a
little more slowly with increasing $N$ than the former.  At the higher
$N$ values typical of Na condensates, this scaling from the simple
model seriously overestimates the output, which begins to saturate
with increasing $N$.

\begin{figure}
\centerline{\epsfxsize=4.25in\epsfbox{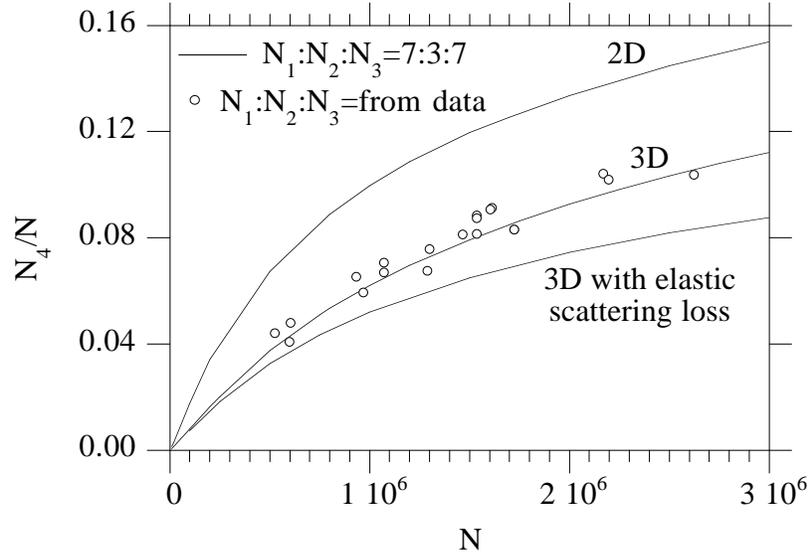}}
\caption {Fraction of atoms in the 4WM output wavepacket, $N_{4}/N$,
versus the total number of initial atoms, $N$, calculated in 2D, 3D
and 3D with inclusion of elastic scattering loss as discussed in
Sec.~{\protect \ref{el_scat}}.  The open circles represent
calculations using experimental data {\protect \cite{Deng}} to
determine the ratios $N_1:N_2:N_3$ rather than taking the nominal
values $N_1:N_2:N_3 = 7:3:7$.  The trap is the same as in
Fig.~\ref{f3}.  The Bragg pulses are applied 600 $\mu$s after the
trapping potential is turned off.}
\label{f12}
\end{figure}

Fig.~\ref{f12} shows three curves giving the fraction of atoms in the
4WM output wavepacket as a function of the initial total number of
atoms $N$ as calculated by (1) 2D-SVEA and (2) 3D-SVEA simulations
without including elastic scattering loss, and as calculated by (3) a
3D-SVEA simulation including elastic scattering loss.  In one set of
calculations we used a ratio of atoms in the three initial wavepackets
of $N_1:N_2:N_3 = 7:3:7$.  These calculations produce the three smooth
curves in Figure \ref{f12}.  In another set of calculations, we used
the measured final fractions from the NIST experiment \cite{Deng} to
determine the initial ratios $N_1:N_2:N_3$, rather than taking the
nominal values $7:3:7$.  The open circles in Figure \ref{f12}, which
no longer fall on a smooth line, show the 3D-SVEA without elastic
scattering for these cases with experimental scatter in initial
conditions.  The relatively small deviation of the points from the
solid curve for the 3D-SVEA without elastic scattering show that the
calculations with the $7:3:7$ ratio is useful for generating a smooth
curve to compare to experimental data.

The effect of including loss from the BEC wavepackets due to elastic
scattering collisions was modeled using
Eqs.~(\ref{SVEA1el})-(\ref{SVEA4el}).  The 4WM output reduction in
Figure \ref{f12} due to elastic scattering ranges from 6 per cent to
16 per cent in going from $10^{5}$ to $10^{6}$ atoms, and becomes more
pronounced for large values of $N$, with the loss due to elastic
scattering reaching 36 per cent for $5\times 10^6$ atoms.  Elastic
scattering of atoms from the different momentum wavepackets removes
atoms from the four BEC wavepackets, and it thereby also lowers the
nonlinear coupling term that gives rise to the 4WM. Although the
mean-free-path for elastic collisions is on the order of 10 times
$r_{TF}$ for $1.5\times 10^6$ atoms, there are a sufficient number of
collisions to make a noticable reduction in the nonlinear output.

\begin{figure}
\centerline{\epsfxsize=4.25in\epsfbox{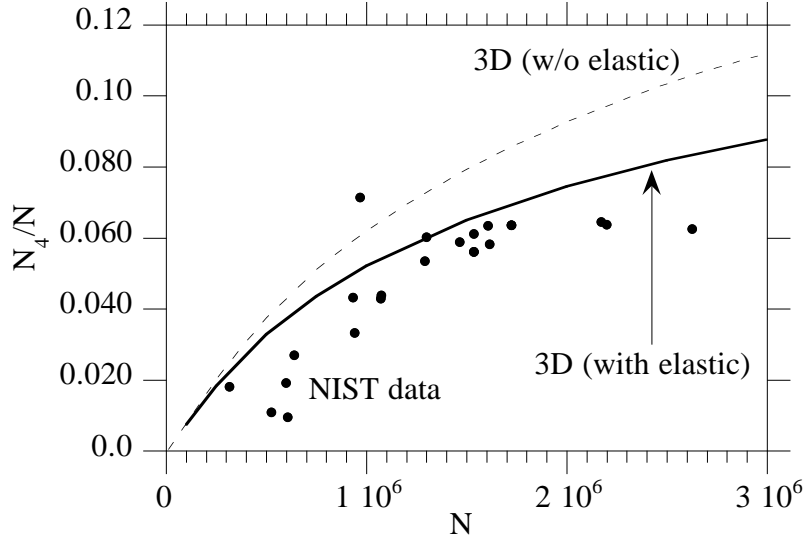}}
\caption {Fraction of atoms in the 4WM output wavepacket, $N_{4}/N$,
versus the total number of initial atoms, $N$, calculated in 3D
without and with inclusion of elastic scattering loss as discussed in
Sec.~{\protect \ref{el_scat}}.  The dots are experimental data
{\protect \cite{Deng}}.  The trap is the same as in Fig.~\ref{f3}. 
The Bragg pulses are applied 600 $\mu$s after the trapping potential
is turned off.}
\label{f13}
\end{figure}

Finally, Fig.~\ref{f13} compares our 3D-SVEA calculation, with
corrections due to elastic scattering, to the observed output 4WM
fraction in the NIST experiment \cite{Deng}.  The overall agreement is
good, given the approximations in the model and the scatter in the
experimental data.  The calculated curve tends to be slightly larger
than the mean of the measured points, and in particular, does not seem
to saturate as fast at large $N$ as the experimental data.  Since
systematic error bars were not given for the data, it is difficult to
know whether this slight disagreement is significant.  There are
clearly approximations in the theory, such as using the GPE method or
ignoring the dynamics during the application of the Bragg pulses. 
There also are effects in the experiment that might have a bearing on
the comparison.  For example, Fig.~2b of reference \cite{Deng}
reported a best case of 10.6 per cent 4WM output for $N=1.7\times
10^6$ atoms, although a lower figure near 6 per cent reported in
Fig.~3 of reference \cite{Deng} was more typical.  The 10.6 per cent
output would disagree with our calculations on the high side.  This
indicates that there is sufficient uncertainty in the quantitative
aspects of the experiment to warrant a more systematic experimental
exploration of the 4WM signal.  Other possible sources of differences
between theory and experiment include micromotion of the initial BEC
in the time-orbiting-trap, laser misalignment, and a small finite
temperature component of the BEC.

\section{Summary and Conclusions and Outlook} \label{conclusions}

We have developed a full description of four-wave mixing (4WM) using a
mean-field treatment of Bose-Einstein condensates.  The
slowly-varying-envelope approximation is a powerful tool that reduces
the numerical grid requirements for calculating the time-dependent
dynamics of fast-moving wavepackets with velocities greater than a
photon recoil velocity.  We find that elastic scattering loss between
atoms in the fast wavepackets removes enough atoms from the
wavepackets to affect the 4WM output.  The quantum mechanical 3D
calculations presented here show good agreement with experiment.

In spite of the strong analogy between atom and optical 4WM, there are
fundamental differences.  In optical 4WM, the energy-momentum
dispersion relation is different than in the massive boson case. 
Because we neither create nor destroy atoms, the only 4WM processes
allowed for matter waves are particle number conserving.  This is not
the case for optical 4WM where, for example, in frequency tripling
three photons are annihilated and one is created.  Particle, energy
and momentum conservation limit all matter 4WM processes to
configurations that can be viewed as degenerate 4WM in an appropriate
moving frame.

We have considered 4WM using condensates of the same internal states. 
The internal states of the atoms can be changed by using Raman
transitions.  Thus, one can envision scattering atoms in one internal
state from the matter-wave grating formed by atoms in a different
internal hyperfine state.  It is also possible to study the details of
4WM between mixed atomic species.  We are in the process of carrying
out such calculations.  Quantum correlations created by the nonlinear
process could lead to the study of non-classical matter-wave fields,
analogous to squeezed and other non-classical states of light.  It is
of interest to investigate such cases.  By varying the magnetic field
to allow a Feshbach resonance to change the $U_0$ coupling parameter,
4WM can be modified dynamically during the dynamics that occur as the
wavepacket fly apart, thus increasing or decreasing 4WM output.  Such
studies are also feasible.

It is possible to modify the mean-field description of 4WM, and more
generally, Bragg scattering of BECs, by generalizing the GP equation
to allow incorporation of momentum dependence of the nonlinear
parameters, thereby putting the treatment of elastic and inelastic
scattering on a firm footing.  This will be presented elsewhere
\cite{BTiesBJ}.

\bigskip

\begin{acknowledgments} This work was supported in part by grants from
the US-Israel Binational Science Foundation, the James Franck Binational
German-Israel Program in Laser-Matter Interaction (YBB) and the U.S.
Office of Naval Research (PSJ).  We are grateful to Eduard Merzlyakov for
assisting with the 3D computations carried out on the Israel Supercomputer
Center Cray computer.  We thank Ed Hagley, Lu Deng, William D. Phillips,
Marya Doery and Keith Burnett for stimulating discussions on the subject.
\end{acknowledgments}

\newpage

\newpage

\end{document}